\documentclass[preprint,aps]{revtex4}
\usepackage[dvips]{graphicx}

\begin{document}

\title{Non-linear and scale-invariant analysis of the Heart Rate Variability}
\author{J.\ Kalda, M.\ S\"akki}
\address{
Institute of Cybernetics,
Tallinn Technical University,
Akadeemia tee 21,
12618 Tallinn,
Estonia}
\author{M.\ Vainu}
\address{
Tallinn Diagnostic Center,
Estonia}
\author{M.\ Laan}
\address{
Tallinn Children Hospital,
Tallinn,
Estonia}
\begin {abstract} 
Human heart rate fluctuates in a complex and non-stationary manner. 
Elaborating efficient and adequate tools for the analysis of such signals
has been a great challenge for the researchers 
during last decades. 
Here, an overview of the main research results in this field is given.
The following question are addressed:
(a) what are the intrinsic features of the heart rate variability signal;
(b) what are the most promising non-linear measures, bearing in mind 
clinical diagnostic and prognostic applications.
\end {abstract} 
\maketitle

\subsection{INTRODUCTION}

The heart rate of healthy subjects fluctuates in a complex manner.
These nonstationary and nonlinear fluctuations are related
mainly to a nonlinear interaction between competing neuroautonomic inputs:
parasympathetic input decreases and sympathetic stimulation
increases the heart rate. Meanwhile, heart pathologies may
decrease the responsiveness of the heart and lead to a failure 
to respond to the external stimuli. Evidently, such pathologies 
lead to an overall reduction of the heart rate variability (HRV).
Understanding the diagnostic and prognostic significance of the 
various measures of HRV has a great importance for 
the cardiology as a whole, because unlike the invasive methods
of diagnostics, the required measurements are low-cost and
are harmless for the patients. 
A particularly important application
is the prognostics of the patients  with increased risk of sudden cardiac death. 
While the ``linear measures'' 
of HRV are nowadays widely used in clinical practice, 
the importance of more complicated measures have been hotly 
disputed in the scientific literature during the recent decades.

The layout of this review is as follows. In the first section, 
general aspects of the heart rate generation, electro cardiogram (ECG)
structure, and data acquisition are discussed.
In the second section, we give a brief overview of the ``linear era'' of the HRV analysis.
Section three is devoted to the early studies of the non-linearity 
of HRV, ie.\ to the methods based on the reconstructed phase space 
analysis. Here we also provide the modern view
on the applicability of these methods. In section four,
we discuss the self-affine and multi-affine aspects of
HRV (including the wavelet-transform-based techniques).
Section five deals with the phenomenon which can be referred to as
``intertwining of low- and high variability periods''.
Section six examines the effect of synchronization between
heart rate and respiration. Section seven provides a brief
conclusion.

\subsection{HEART RATE GENERATION, ECG,  AND DATA ACQUISITION}

The quasi-periodic contraction of cardiac muscle is governed 
by the electrical signal, which is generated by the sino-atrial (SA) node --- 
a set of electrically active cells in a small area of the right atrium.
The signal spreads through the atrial muscle leading to its contraction. 
It also spreads into a set of specialized cells - the atrio-ventricular (AV) node. 
Further the signal spreads via the His-Purkinje bundle (which is a 
fractal-like set of electrically conductive fibers) to the myocardial cells 
causing their contraction. ECG is measured as the electrical potential between 
different points at the body surface. The activity of the SA node by itself is 
not reflected on the ECG. The electrical activation of the atrial cells leads to the 
appearance of the P-wave of the ECG. Q, R, S and T waves (see Fig.~1) 
are caused by the electrical activity of the 
ventricular muscle. The heart rate is generally measured as the 
RR-interval $t_{RR}$ --- the time-lag between two 
subsequent R-pikes (R-pike itself corresponds to the ventricular contraction). 
For the HRV analysis, only the normal heart activity is taken into account. 
All the QRS-complexes are labelled as normal or arrhythmic. 
Note that even for healthy patients, some heartbeats can be arrhythmic. Normal-to-normal 
interval $t_{NN}$ is defined as the value of $t_{RR}$ for such 
heartbeats, which have both starting and ending R-pikes labelled as normal (see Fig.\ 1). 
\begin{figure}[hb]
\hfill
\includegraphics{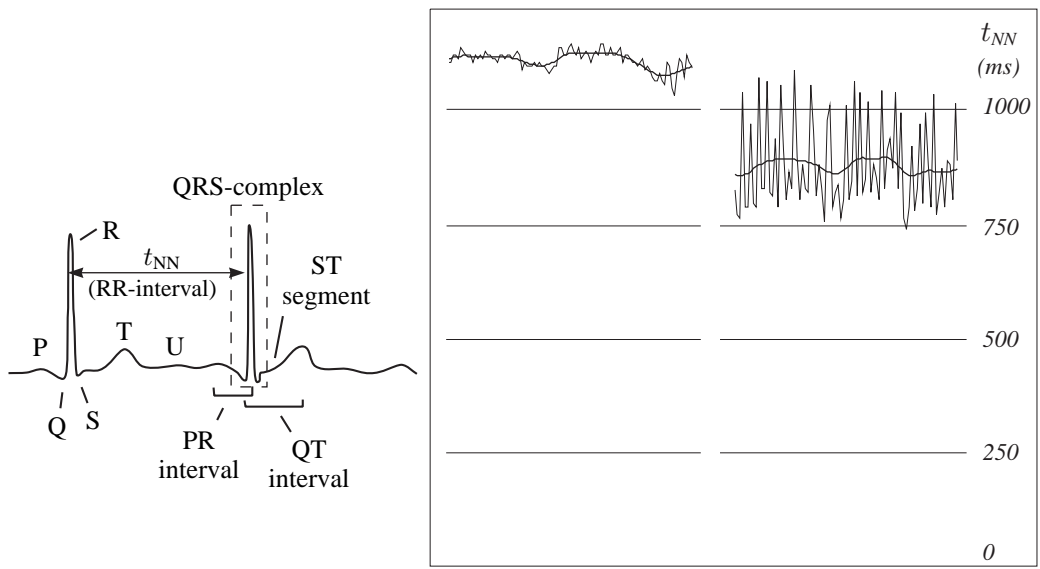}\hfill~
\caption[]{Left image: normal ECG recording. Image at right: $t_{NN}$ sequences of low- and high variability.}
\end{figure}

Typically, HRV analysis is based on the 24-hour recordings of the {\em Holter-monitoring}. Shorter 
ECG recordings can be used for this purpose, as well; however, in that case it is impossible to 
observe the long-scale variations and compare the sleep-awake differences in the heart rhythm. 
Portable apparatus stores the ECG data as the time-dependent voltage $U(t)$ either on a tape or on a PC 
flash card; the sampling rate is 125 Hz or higher. 
The data are later analyzed by computer software. Typical commercial 
software allows visualization of the ECG recording, automated or 
semi-automated recognition of arrhythmias and artifacts, and the calculation 
of the standard ``linear'' characteristics of the HRV. 
Most often, a research devoted to the methods 
of non-linear dynamics is based on plain sequences of NN-intervals, 
and disregards the details of the continuous ECG recordings.
Other aspects of ECG, eg.\ the clustering of arrhythmic beats \cite {Liebovitch} and 
dynamics of QT intervals \cite {Lass} are also of high clinical importance, but somewhat understudied
and will be  not discussed here.
\begin{figure}[hbt]
\hfill
\includegraphics{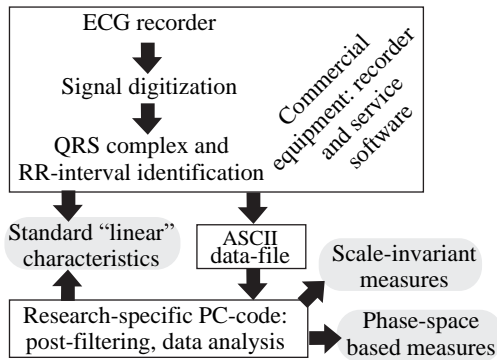}\hfill~
\caption[]{The analysis of the heart rate variability: the scheme of data acquisition and analysis.}
\end{figure}

The experimental data serving as the bases of the original research 
performed by the authors of the review
were recorded {\em (a)} at the Tallinn N\~ omme Hospital (children)
and {\em (b)} Tallinn Diagnostic Centre (adult subjects);
the scheme of data acquisition is presented in Fig.\ 2.
For the group {\em (a)}, the recordings of ambulatory Holter-monitoring covered 12 healthy subjects of
mean age $11.5\pm 3.3$ years, 6 children with clinically documented sinus node disease  
(mean age $11.5\pm 1.9$ years), and 12 subjects with miscellaneous diagnosis. 
The sampling rate of the ECG was 125Hz. 
For the group {\em (b)}, specifics are given in Table~1. These data have been 
obtained during regular diagnostical examinations of more than 200 patients, the 
ECG sampling rate has been 180 Hz.
The diagnostics and data verification has been made by qualified cardiologist.
The data preprocessing included filtering out falsely detected QRS-complexes (artifacts and arrhythmias).
\begin {table}[!bth]
\begin {tabular}{|l|c|c|c|c|c|c|c|}
\hline
         & Healthy & IHD & SND  & VES & PCI & RR & FSK\\
 \hline
No.\ of patients & 103    & 8   & 11   & 16  & 7   & 11 & 6 \\
\hline
Mean age & $45.5$ & $65.4$ & $50.0$ & $55.9$ & $47.3$ & $55.5$ & $11.7$\\
\hline
Std.\ dev.\ of age & $20.5$ &  $11.4$ & $19.3$ &  $14.3$ &  $11.6$ &  $14.4$ &  $4.6$ \\
\hline
\end{tabular}
\caption{
Test groups of patients. Abbreviations are as follows: IHD - Ischemic Heart Disease (Stenocardia); SND - Sinus Node Disease; 
VES - Ventricular Extrasystole; PCI - Post Cardiac Infarction; RR - Blood Pressure Disease; 
FSK - Functional Disease of Sinus Node.}
\end{table}

\subsection{LINEAR MEASURES OF HRV}

The clinical importance of the heart rate variability (HRV)  was first
noted in 1965 by Hon and Lee \cite{Hon}. 
Since then, the statistical properties of the
interbeat interval sequences have attracted the attention of a
wide scientific community. An increased risk of post-infarction
mortality was associated with the reduced HRV by Wolf
et al.\ in 1977 \cite{Wolf}.

Wider attention to the problem has been attained in the early 1980s, when
Akselrod et al.\ introduced the spectral methods for the
HRV analysis  \cite{Aksel}. The spectral characteristics are generally referred to as 
``frequency-domain characteristics'' and are opposed to the 
``time domain methods'', which are derived directly from
the $t_{NN}$-sequence. In the late 1980s, the clinical importance of HRV
became generally recognized. Several studies confirmed that HRV was a strong and independent
predictor of mortality following an acute myocardial
infarction \cite{Kleiger}--\cite{Bigger}.
As a result of this, a breakthrough has been achieved:
the ``linear'' measures of HRV  became important tools of clinical practice.

A non-exhaustive list of the parameters which are currently used in medical practice, is as follows: 
the mean NN interval;
the difference between night and day heart rate; longest and shortest NN intervals; the
standard deviation of the NN interval (SDNN, typically calculated over 24-hour period);
the standard deviation of locally (usually 5 min) averaged NN intervals (SDANN); the
mean of the 5-minute standard deviation of the NN interval (averaged over 24 h; SDNN
index); the square root of the mean squared differences of successive NN intervals
(RMSSD), the percentage of interval differences of successive NN intervals greater than
50 ms (pNN50), the spectral power of high- and low-frequency fluctuations in
NN-sequences.

\subsection{RECONSTRUCTED PHASE SPACE}

It is widely accepted that the heart rhythm generation in the complex of sinus-node
and atrio-ventricular node can be well described by nonlinear dynamical models, 
where SA node and AV node form a system of nonlinear coupled oscillators \cite{West,JE}.
The model has been proven to be viable
and predicts several experimentally observed phenomena, such as
Wenckebach and Mobitz type II arrhythmias and bistable behavior \cite{JE}.  
This deterministic non-linear model predicts that the phase trajectories of an healthy heart lie on an attractor of the 
coupled system of oscillators. Consequently, one should be able to observe well-defined patterns on the Poincar\`e sections of the 
phase-space.  Note that in the case of physiological data, there is no information, what might be the canonical variables. 
Therefore, the phase trajectory is reconstructed in time-delay coordinates $U(t)$, $U(t+\tau)$, \dots, $U[t+(D-1)\tau]$ 
[or $t_{NN}(n)$, $t(n+1)$, \dots, $t(n+D-1)$]. Here $D$ is the so called embedding dimensionality, i.e. the dimensionality 
of the reconstructed phase-space. It is expected that the real phase trajectory is mapped to the  reconstructed 
trajectory by a smooth transform.

Exactly such a reasoning has lead to the idea that the dynamical characteristics from the theory of
non-linear dynamics could be used for the diagnostic purposes.  The early studies by Babloyantz et al.\ 
\cite{Bablo} gave rise to extensive studies in 1990s \cite{Poon}--\cite{Govind}.   
The experimental observations seemingly confirmed the theoretical expectations.
Particularly, the correlation dimension of the continuous ECG recording (i.e. the recorded voltage 
as a function of time) has been reported to be between 3.6 and 5.2.
The conclusion has been that the dynamics  of the heart of healthy persons is 
less regular than that of persons with severe cardiac pathologies. Correspondingly, the correlation dimension 
has been often thought to be a measure for the healthiness of the heart. 
The other tools of the analysis of non-linear dynamical systems (such as Lyapunov exponents; 
Kolmogorov, Shannon, pattern, and approximate entropies; etc) have been exploited in an equal extent.

The correlation dimension of a data sequence is typically calculated according 
to the Grassberger-Procaccia algorithm \cite{Grassb}.
In a reconstructed phase space of dimensionality $D$, the correlation 
sum $C=\sum_{i,j}\theta(r-|\bm r_i - \bm r_j|)$ is calculated as a function of the radius $r$; 
it is expected to behave as a power-law $C \propto r^{\nu(D)}$. 
Here $\bm r_i$ denotes the $D$-dimensional radius-vector of the $i$-th data-point, and 
$\theta(r)$ stands for the Heaviside function.
The correlation dimension $d_c$ is found as the limit of $\nu$ at large
values of $D$ (in fact, it is expected that for $D>d_c$, 
the exponent $\nu$ is independent of $D$, and in that case $\nu=d_c$).

However, there are various arguments leading us to the conclusion that
the formally calculated correlation dimension of a heart rhythm does not correspond to the dimensionality of 
an intrinsic attractor; similarly, the formally calculated Lyapunov exponents, entropies etc.\ 
do not describe the respective aspects of underlying non-linear dynamics.
{\em First,}
it has been pointed out that physiological time-series are 
typically non-stationary and noisy, and therefore, the correlation 
dimension cannot be calculated reliably \cite {Kantz,Kanters,Bezerianos}; this fact is
nowadays widely accepted.
In the case of human heart,
the ``noise'' comes from the autonomous nervous system in the form of inputs regulating the 
heart rate (cf.\ \cite {Berne,Kaplan,Rosenblum}):
from the viewpoint of underlying nonlinear deterministic system, these 
effectively  non-deterministic signals perform the role of high level noise.
It should be also noted that some inputs of the autonomous nervous system
may lead to quasi-periodic signals --- an easy source of false detection of
low-dimensional chaos and apparent patterns in simple time delay maps, see Fig.~3--4.
Thus, respiration gives rise to the signal of typical period of 4s; 
the effect is most pronounced when the patient is  at rest, and is stronger for young persons.
{\em Second,} it has been emphasized that a reasonable fitting of a correlation sum to a power law
does not necessarily mean that the obtained exponent is the correlation dimension
of the underlying dynamical system; instead, thorough non-automatable verification procedure has 
to be done \cite {KantzTextbook}.
{\em Third,} the length of the data sequences is often inadequate for reliable 
calculation of high values of the correlation dimension $d_c \agt 6$, cf.\  
\cite{KantzTextbook,Govind}.

\begin{figure}[h]
\hfill
\includegraphics{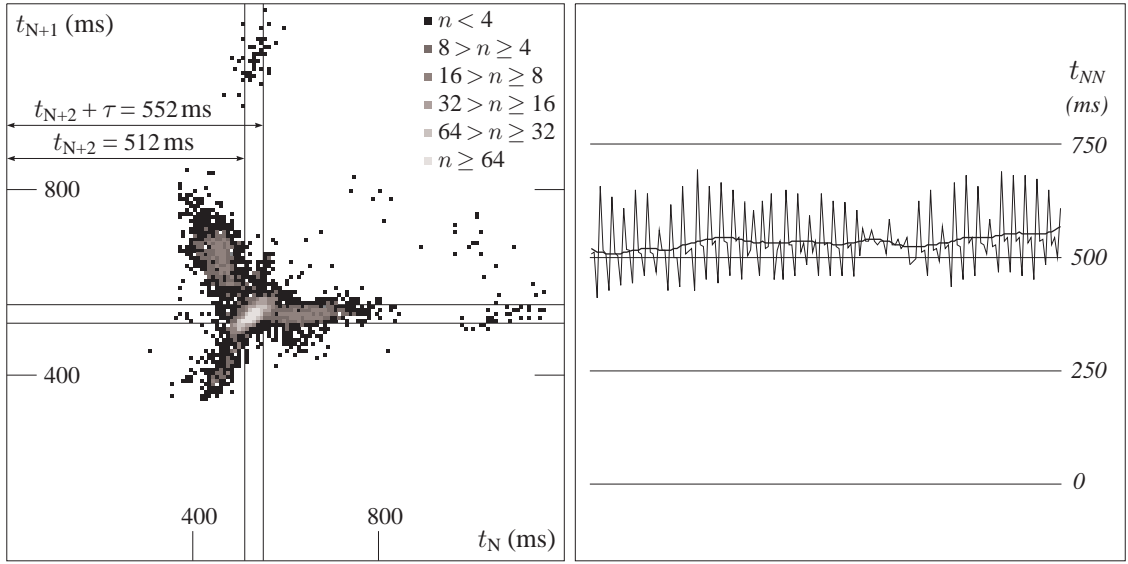}\hfill~
\caption[]{A cross-section of the 3-dimensional reconstructed phase space for a patient with pronounced 4:1 mode-locking 
(see also Section G); around the central cloud of points, 
three major satellite-clouds can be seen; these satellite-clouds correspond to the sequence 
of interbeat intervals, shown on the right-hand plot. The observed oscillations with period 4 can be attributed to 
the modulation of the heart rate by respiration.}
\end{figure}
\begin{figure}[h]
\hfill
\includegraphics{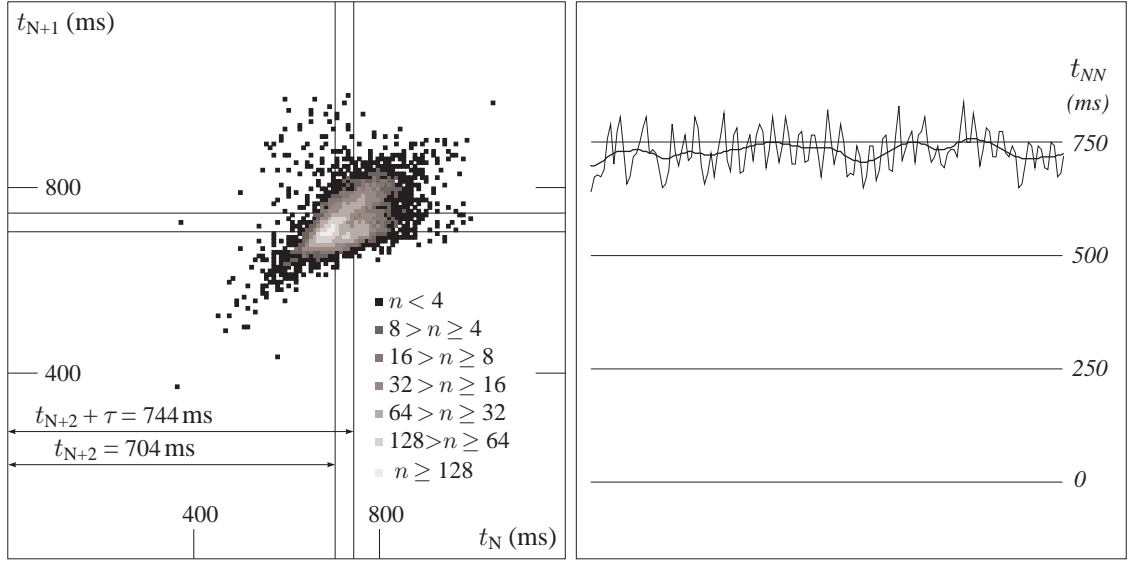}\hfill~
\caption[]{The same as in Figure 3. Mode-locking (4:1 and 5:1) is weaker, but the heart rate 
modulation by the respiration is significant.
One can distinguish
two branches of the central cloud, which are caused by the respiratory modulation.}
\end{figure}

The above discussed research results can be summarized as follows:
{\em (a)} the correlation sums of human heart rate follow typically a scaling law;
{\em (b)} in most cases, the scaling exponents are not the correlation dimensions.
This leads us to a natural question: what is the physical meaning
of these formally calculated exponents?
Our answer to this question is based on simple observations, valid for
healthy patients: {\em(a)} the 
long-time variability of the inter-beat intervals is
typically much higher than the variability on the time scale of few heart beats;
{\em(b)} for those periods, when the mean heart rate is high (when the subject is
performing physical exercise) the heart rate variability is low;
{\em(c)} the heart rate is controlled  by effectively random non-deterministic 
inputs arriving from the autonomous nervous system.
As a consequence, in time delay coordinates, an HRV time-series generates a 
baseball bat-shaped cloud of points.
Although the theoretical value of the correlation dimension of such a cloud is infinite, the
finite resolution of the recording apparatus, finite length of the time-series, and the linear 
structure of the cloud result in a  smaller value. This is evident for a very narrow ``bat'', 
which is efficiently one-dimensional. 

Our conjecture passes also a quantitative test: the correlation sum of
surrogate data-sets constructed using Gaussian random data-series and mimicking the features {\em (a)--(c)} 
(see Fig.~5) scales almost identically to that of clinical HRV data, see Fig. 6 and Ref.~\cite{MS}.
\begin{figure}[h]
\hfill
\includegraphics{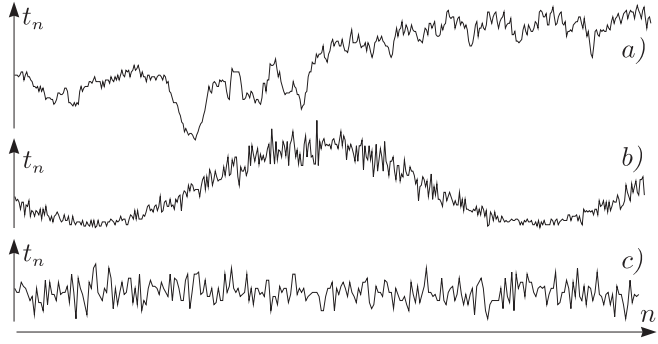}\hfill~
\caption[]{Time series for real HRV data {\em (a)}, surrogate data {\em (b)}, and Gaussian noise {\em (c)};
the beat interval $t_n$ is plotted versus the beat number $n$.}
\end{figure}
\begin{figure}[h]
\hfill
\includegraphics{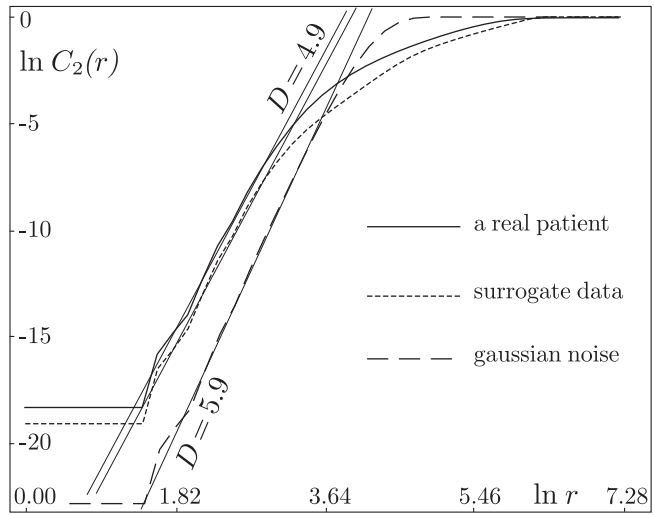}\hfill~
\caption[]{The correlation sum $C_2(r)$ (as a function of the radius $r$) of surrogate data scales 
almost identically to the real clinical data.}
\end{figure}

To conclude, the reconstructed-phase-space-based measures fail in describing
a deterministic chaos inside the heart, because the deterministic dynamics is suppressed
by essentially intermittent signals arriving from the autonomous nervous system and
regulating the heart rhythm. However, some fine-tuned measures (eg.\ various entropies, cf.\  \cite{Zebr})
can be useful in describing the level of short-time variability of the heart rhythm, and
complement the linear quantity pNN50 (which also measures the high-frequency component of HRV).

\subsection{SCALE-INDEPENDENT MEASURES.}

Recent studies have shown that scale-invariant characteristics can be
successfully applied to the analysis of the heart rate variability \cite{Peng}--\cite{Amaral}. 
However, this conclusion has been disputed, and certain scale-dependent measures
(particularly, the amplitude of the wavelet spectra at specific time-scale) 
have been claimed to provide better results \cite{Thurn}.
The scale-independent methods have been believed to be more universal, subject-independent, 
and to reflect directly the dynamics of the underlying system, unlike 
the scale-dependent methods which may reflect characteristics specific to
the subject and/or to the method of analysis  \cite{Amaral}.
Opposing argument has been that certain heart disorders affect the heart rate
variability at a specific scale or range of scales; owing to this circumstance, 
at the properly chosen time-scale, scale-dependent measures may provide a useful information \cite{Thurn}. 

The simplest relevant scale-independent measure is the Hurst exponent $H$, 
which has been introduced to describe statistically self-affine random 
functions $f(\bm r)$ of one or more variables \cite{Mand}.  Such a  function is referred to as 
{\em fractional Brownian function} and  satisfies the scaling law 
$$\left<[f(\bm r_1) -f(\bm r_2)]^2 \right> \propto |\bm r_1 - \bm r_2|^{2H}.$$
Note that $H=1/2$ is a special case of ordinary Brownian function ---
the increments of the function are delta-correlated, and $f(r)$ can be thought to be the displacement of a 
Brownian particle as a function of time $r$. Therefore, in the case of $H<1/2$,
there is a negative {\em long-range correlation} between the increments of the function. Analogously, $H>1/2$ 
corresponds to a positive correlation. 
Note that the early scale-invariant studies of HRV were based on power spectra \cite{Kobay,Yama}, an aspect 
closely related to the scaling exponent $H$.

Many phenomena in nature exhibit this kind of scale-invariance, and lead to fractional Brownian 
time-series \cite{Mand}. The same is true for the heart rate variability: after filtering out 
short-scale components  with $\tau < 30$s (corresponding to the respiratory rhythm, to the blood-pressure oscillations, 
and to the pathological Cheyne-Stokes respiration), the fluctuation function $F(n)$, defined as
\begin{equation}
F(\nu ) = \left< |t_n - t_{n+\nu}|\right>	
\end{equation}
revealed a good 
scaling behavior $F(\nu)\propto \nu^H$ \cite{Peng}. While for healthy patients, 
the increments of the heart rhythm were found to be significantly anticorrelated resulting in $H<1/2$, 
the heart rhythm of the patients with dilated cardiomyopathy
was essentially Brownian with $H \approx 1/2$ \cite{Peng}. 
In the case of our patient groups, there was no significant correlation between the diagnosis and the Hurst exponent,
and there were many  healthy subjects with  $H \approx 1/2$, see Fig.~7.
Finally, various techniques, such as detrended fluctuation analysis \cite{Peng2},  
detrended time series analysis \cite{Ashkenazy2}, and wavelet amplitude analysis \cite{Ashkenazy} have been 
proposed to fine-tune the Hurst-exponent-based approach.

\begin{figure}[h]
\hfill
\includegraphics{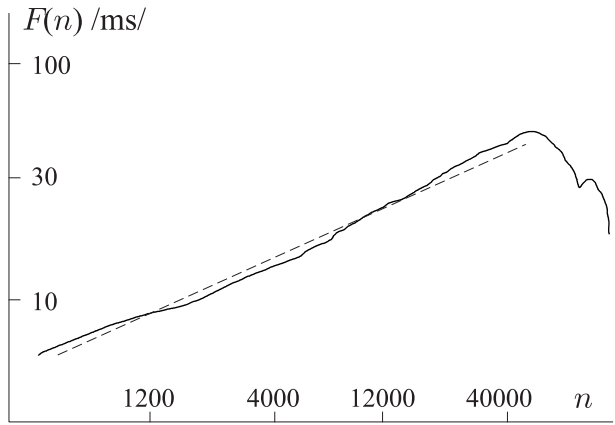}\hfill~
\caption[]{The fluctuation function $F(\nu)$ is plotted versus the time lag $\nu$.
Almost straight line indicates a good scaling behavior $F(\nu)\propto \nu^H$ (here with $H=0.50$).}
\end{figure}

Complex non-stationary time-series cannot be described by a single scaling exponent $H$. Indeed,
simple scaling behavior is expected if there is a Gaussian distribution of increments. However,
even in the case of Gaussian functions, the scaling exponent is not necessarily constant over the whole range
of scales. Instead, it can be a slow (eg.\ logarithmic) function of the scale, so that
other descriptions (such as stretched exponentials) may be required.
Physiological time-series are typically non-Gaussian. For such functions,
scale-invariance can be very complicated. A non-exhaustive way to describe
such behavior is to calculate the  multifractal spectrum of Hurst exponents \cite{Mandel}.
Therefore, it is not surprising that the human heart rate signal was found to obey a 
multi-affine structure \cite{Amaral,Ivan}. 

Qualitatively, a multifractal time-series behaves as follows. 
If the whole time-series is divided into short segments, each segment can be characterized by its own Hurst exponent $h$
(referred to as the Lipschitz-H\" older exponent). 
Then, the distribution of segments of fixed values of $h$ is self-similar, and is described by a
fractal dimension $f(h)$. Technically, the spectrum $f(h)$ can be calculated 
by the means of wavelet transform, cf.\ \cite{Amaral}. This scheme includes the calculation
of the scaling exponents $\tau(q)$ (referred to as the mass exponents), which 
describe, how the $q$-th moment  of the wavelet transform amplitude 
scales with the wavelet width. The scaling exponents $\tau(2)$ and $\tau(5)$ have been found to 
have a significant prognostic value (for the post-infarction prognosis) \cite{Amaral}.
The wavelet transform amplitudes, calculated for a specific wavelet width ($\approx 5$ min)
have been claimed to be of even higher prognostic value \cite{Thurn}. However, 
independent studies have shown that the scale-invariant measures seem to be 
superior tools \cite{Saermark}. It should be also noted that the wavelet transform amplitude at
a fixed time-scale is closely related to the linear measure SDANN.
Substituting the robust standard deviation by a wavelet transform amplitude is a
technical fine-tuning which cannot be expected to result in a qualitatively new information.

The multifractal structure of the heart rate signal has several consequences. Thus, the 
$q$-th order structure function (a concept borrowed from the theory of the fully-developed turbulence) 
of the heart rate interval has  a scaling behavior, with the scaling
exponent $\zeta(q)$ being a function of $q$\cite{Lin}. Note that this spectrum of exponents is
very closely related to the above-mentioned $\tau(q)$ spectrum (both describing the same physical phenomenon,
differences being of a technical kind).
However, the wavelet-transform-based technique makes a more complete utilization of the underlying data
and  therefore,  $\tau(q)$ spectrum can be expected to yield somewhat superior prognostic and/or diagnostic results.

Another aspect related to the multifractal nature of the heart rhythm, is the multiscale entropy (MSE) \cite{Costa}.
While the single-scale entropies (approximate entropy, Shannon entropy) are related to the
short-time dynamics of the heart rhythm and to the probability distribution function of points 
in the reconstructed phase space, the multiscale entropy extends these concepts to longer 
time-scales. MSE is not directly reducible to the multifractal spectra $f(h)$ [or $\tau(q)$]; however, 
both techniques address the question of how wide is the range of dynamics for the mean heart rate 
(averaged over a time $T$), depending on the time-scale $T$. The clinical usefulness of the 
MSE is still unclear (apart from the fact that it has been claimed to distinguish between
healthy subjects and patients with congestive heart failure \cite{Costa})

\subsection{INTERMITTENCY OF THE HRV}

Multifractal spectrum addresses only one aspect of the non-Gaussianity of the time-series increments
by revealing the possible range of scaling laws for the long-range [at time-scale of many ($\gg 1$) 
heartbeat intervals] dynamics of mean heart rhythm. However, the short-time variability of heart rhythm 
is also fluctuating in a complex manner.
It has been pointed out that the NN-sequences of healthy subjects consist of intertwined high- and
low-variability periods \cite{Poon}. This conclusion can be easily
verified  by a simple visual observation of the NN-sequences, see Fig.\ 8. 
The quantitative analysis of such a behavior is based on the distribution law of the low-variability 
periods \cite{JK0,JK}, which will be discussed below. 
Another aspect of such an intertwining is the clustering of the periods of similar mean heart rate:
the heart rate signal can be divided into segments of different
mean heart rate, with distinct boundaries between these segments; 
there is a power-law segment-length distribution 
of the segments \cite{Galvan}.

\begin{figure}[h]
\hfill
\includegraphics{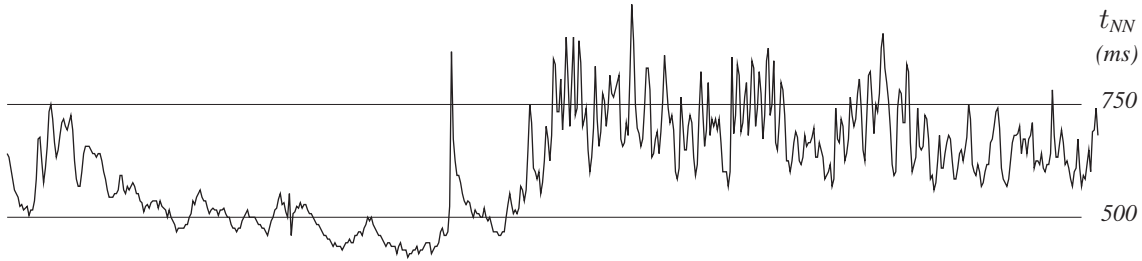}\hfill~
\caption[]{For healthy patients, the high- and
low-variability periods of the heart rhythm are intertwined.}
\end{figure}

In order to analyze quantitatively the intertwining of high- and
low-variability periods, we have studied distribution of low-variability periods and showed
that typically, it follows a multi-scaling Zipf's law.
Originally, the Zipf's law has been formulated by G.\ K.\ Zipf for the frequency of words in 
natural languages \cite{Zipf}. For a given language (e.g.\ English), the frequency (the number of occurrences 
divided by the total number of words) of each word is calculated on the bases of 
a large set of texts. The ranks are determined by arranging the words according to their frequency $f$:
the most frequent word obtains rank $r=1$, the second frequent --- $r=2$ etc. 
It turns out that for a wide range of ranks (starting with $r=1$), there is a power law $p(r) \propto r^{-\alpha}$, 
where $\alpha \approx 1$. This law is universal, it holds for all the natural languages 
and for a wide variety of texts \cite{Zipf}. Furthermore, similar scaling laws describe the 
rank-distribution of many other classes of objects, as well. Thus, when cities are arranged according 
to their population $s$, the population of a city $s \propto r^{-\alpha}$, with $\alpha \approx 1$ \cite{Zipf}.
Another example is the income-rank relationship for companies; here we have again  $\alpha \approx 1$ \cite{Zipf}.
In the most general form, the law can be formulated as $p \propto (r+r_0)^{-\alpha}$, and $\alpha$ is 
not necessarily close to unity \cite{Mandel}. This more general form of the law can be applied to the distribution of scientists 
according to their citation index, to the distribution of internet sites according to the 
number of visitors etc.

The Zipf's law is characteristic to
such dynamical systems at statistical equilibrium, which satisfy the following conditions: 
{\em (a)} the system consists of elements of different size; 
{\em (b)} the element size has upper and lower bounds; 
{\em (c)} there is no intermediate  intrinsic size for the elements.
The human heart rate, when divided into the low-variability periods,
satisfies all these requirements.
The duration $\tau$ of these periods varies in a wide range of scales, 
from few to several hundreds of heart beats.
Thus, one can expect that the rank-length distribution $r(\tau)$ follows the Zipf's law,
\begin {equation}
r \propto \tau ^{-\gamma}.
\end {equation}

First we have to define the local heart rate variability as the
deviation of the heart rate from the local average, 
\[\delta(n)=[t_{NN}(n)-\left<t_{nn}(n)\right>]/\left<t_{nn}(n)\right>;\]
the local average is calculated using a narrow ($\approx 5$-second-wide) Gaussian weight-function.
Then, the low-variability regions are defined as consecutive sequences of
intervals with $|\delta(n)|<\delta_0$; the length $\tau$ of such a region is measured as the number of beats in the sequence.
Further, all the low-variability regions are numbered (to identify them later), and arranged according to 
their length; regions of equal length are ordered randomly.
In such a way, the longest observed region obtains rank $r=1$, second longest --- $r=2$, etc.
Typically, the length-rank relationship reveals multiscaling properties, i.e.\, 
within a certain range of scales, the scaling law (2) is observed,
the scaling exponent $\gamma$ being a (non-constant) 
function of the threshold level, $\gamma = \gamma(\delta_0)$; see Fig.\ 9.

\begin{figure}[h]
\hfill
\includegraphics{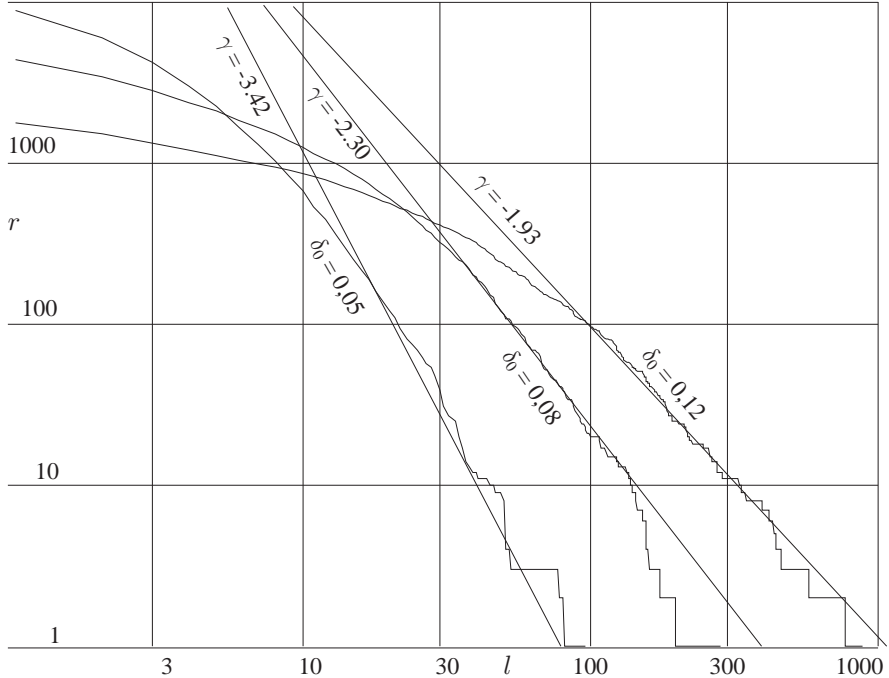}\hfill~
\caption[]{Multi-scaling behavior: the rank of low-variability intervals is plotted against 
the length of the intervals. The scaling exponent $\gamma$ depends on the threshold value $\delta_0$.}
\end{figure}

It is not surprising that the scaling behavior is not perfect.
Indeed, the heart rhythm is a non-stationary signal affected by the non-reproducible daily 
activities of the subjects. The non-stationary pattern of these activities, together with their time-scales,
is directly reflected in the rank-length law.
This distribution law can also have a fingerprint of the characteristic time-scale (10 to 20 seconds) 
of the blood pressure oscillations (which modulate the level of HRV, cf.\ \cite{Kurths}). 
It should be emphasized that the problem of the non-reproducible daily activities 
affects also the reliability of the other scale-invariant measures and is probably the main
obstacle preventing the clinical application of the seemingly extremely efficient diagnostic 
and prognostic techniques.
Finally, there is a generic reason why the Zipf's law is non-perfect at small rank numbers:
while the Zipf's law is a statistical law, each rank-length curve is 
based only on a single measurement. In particular, there is only one longest low-variability period (likewise, only one 
most-frequent word), the length of which is just as long as it happens to be, there is no averaging whatsoever.
For large ranks, the relative statistical uncertainty can be estimated as $1/\sqrt{r}$.

The distribution function of the low-variability periods as a whole 
contains a significant amount of diagnostically valuable information, which is not covered by 
any other (linear or nonlinear) measure of HRV.
The most part of this information seems to be reflected (according to the Student test analysis using the test groups
of Table 1)  by the parameters $r_{100}$ (the rank of the interval with $\tau=100$), $r_{\max}$
(the maximal observed rank), and $\tau_{\mbox{\scriptsize end}}$ (the scale at which the scaling law breaks;
for a precise definition, see Ref. \cite{JK}). These measures allow a clear distinction between 
the healthy subjects and the IHD, VES, and PCI groups \cite{JK}, see also Table 2.
\begin{table}[!tb]
\includegraphics{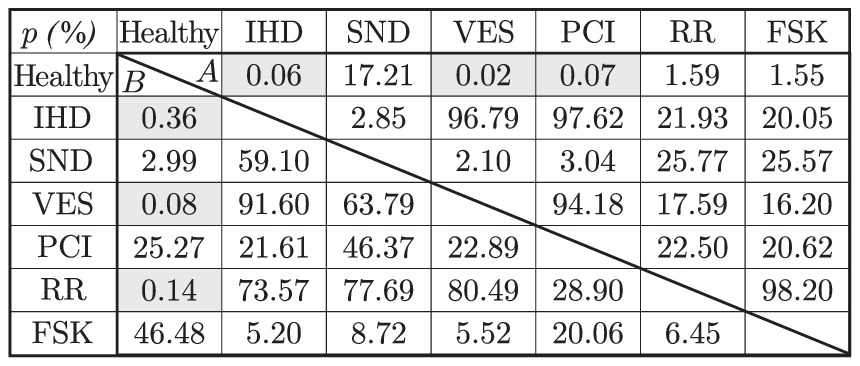}
\caption{$p$-values of the Student test.
Data in the topmost triangular region (with label {\em A}) are calculated using 
the parameter $\ln \tau_{\mbox{\scriptsize end}}$. Triangular region {\em B} corresponds to the parameter 
$\ln r_{\max}$. 
Since multiple tests were carried out,  modified Bonferroni correction \cite{Jaccard} has to be  applied.
Gray background highlights the tests with the adjusted significance $p^\prime < 3\%$.
The control parameter value $\delta_0=0.05$ has been used.}
\end{table}

\subsection{MODE LOCKING BETWEEN HEART RHYTHM AND RESPIRATION}

As mentioned above, respiration affects (modulates) the heart rhythm. This effect is 
mediated by the blood pressure, and the effect known as baroreflex (heart rhythm depends on the 
blood pressure). The heart is most responsive  with respect to the signals of the autonomous nervous
system when the heart rate is slow, i.e.\  when the patient is at rest.
In that case, the heart rate variability is driven by weaker signals, like 
the signals induced by respiration, which (due to their quasi-periodic nature)
may lead to a mode-locking. In the case of mode-locking, the 
heart rate is automatically slightly adjusted so that the respiration and
heart beat periods relate to each other as (small) integers. As a result,
the decorrelation time between the heart rhythm and respiration can be very long.
This is the effect which is in most cases the cause of the patterns (isolated clouds of points) 
observable in the reconstructed phase space (see Fig.~3).

The mode-locking has been studied using bivariate data (simultanious ECG and respiration data)
and the technique called  cardiorespiratory synchrogram \cite{Kurths}. 
Also, a univariate data analysis method using the  angle-of-returntime map has been elaborated \cite{Janson}.
In that case,  the data-set is used to reconstruct the phase of forcing (breathing) and the phase of
oscillator (heart). These phases are plotted versus each other; in the case
of mode-locking, disjoint clouds of points will appear.

Recently, we have developed an 
independent, intuitive and easy to use method of mode-locking detection
from univariate data (RR-interval sequence), which is based on
analysis of the fluctuation function $F(\nu)$, defined by Eq.~1 \cite{MS}.
The fluctuation function of the patients with mode-locking
revealed a presence of an oscillatory component, see Fig.~10b. By dividing the
entire 24-hour HRV record into one-hour intervals, and calculating the amplitude of the
oscillatory component (via a wavelet transform) of the fluctuation function for each interval, we were able to
locate the periods responsible for the satellite clouds in the reconstructed
phase space. These were always the periods before falling asleep, 
around 10 or 11 pm, characterized by a low heart rate and a high respiration-driven
short-time variability. The phase between the heart rate and respiration is 
locked during tens of seconds, confirming the observations of Kurths et al.\ \cite{Kurths}.
Thus, in a certain sense, the heart and respiratory complex act as a system of coupled 
oscillators. 
Finally we note that a specific feature of the patients with strong mode-locking 
was the presence of well-defined ``satellite clouds'' in time-delay map (see Fig.~3).
Therefore, the time-delay map can be also used to detect mode-locking; however,
this method is non-quantitative, less sensitive than the fluctuation-function-based
technique, and does not give a hint, which mode-locking modes are observed. 
The presence of  a natural quantitative measure [the wavelet transform amplitudes]
is also the main advantage of our approach  over the alternative method.

As compared with the alternative techniques,
our method of mode-locking detection is very simple and does not require synchronous
respiration rhythm recording (unlike the thorough method \cite{Kurths}), and 
can be conveniently used to find relatively short ($\agt$ 10\,min) locking periods 
from a 24-hour recording. Besides, it provides a natural measure to quantify the degree
of mode-locking (unlike the method of using the  angle-of-returntime map \cite{Janson}).

\begin {figure}[!bth]
\includegraphics*{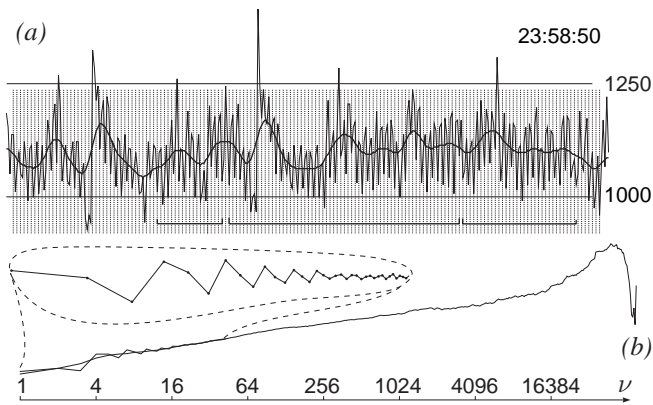}
\caption{Patient with 3:1 mode locking between heart rate and respiration:
{\em (a)} heart beat intervals (in milliseconds) plotted versus the beat number.
Heart rate has a pronounced oscillatory component;
vertical lines mark the period of three heart beats,  horizontal lines indicate the 
sequences  with coherent phase. {\em (b)}
Fluctuation function (arbitrary units) is plotted versus the time lag $\nu$ (in heart beats); the 
oscillating component is magnified.}
\end{figure}

\subsection{CONCLUSIONS}

Below is an attempt to 
classify the measures of heart rate variability. 
\begin{enumerate}
\item ``linear''  methods --- based on standard statistical measures and on the Fourier analysis. 
These are the only methods, which are widely used in clinical practice.
\item ``nonlinear'' methods:
\begin{enumerate}
\item scale-invariant methods:
\begin{enumerate}
\item single-scaling analysis (calculation of the Hurst exponent $H$);
\item multi-scaling analysis --- calculation of the exponent spectra [Lipschitz-H\" older
spectrum $f(h)$, mass exponents $\tau(q)$, or structure function exponent spectrum $\zeta(q)$];
these seem to be the most promising measures, at least for prognostic purposes;
\item calculation of the multi-scale entropy;
\item analysis of the HRV-data segments with similar mean heart rate;
\item analysis of the distribution law of low-variability periods (performs well in diagnostic tests,
there are no prognostic tests yet);
\end{enumerate}
\item scale-dependant methods:
\begin{enumerate}
\item performing a phase-space analysis (entropy-based measures, correlation dimension, Lyapunov exponents etc);
\item calculating wavelet spectra at specific time-scales;
\item heart rhythm and respiration mode-locking analysis.
\end{enumerate}
\end{enumerate}
\end{enumerate}

Human heart rate fluctuates in a complex and non-stationary manner. 
Elaborating efficient and adequate tools for the analysis of such signals
has been a great challenge for the researchers 
during last decades. 
The above given long list of nonlinear techniques proves that 
the research has been successful,
various important features of such time-series have revealed.
However, there is no consensus of which methods are most efficient 
from the point of view of clinical applications.
On the one hand, this is caused by the high non-stationarity
and irreproducibility of these time-series: the complex measures
of HRV depend not only on the healthiness of the heart, but also 
on the daily habits of the subject \cite{Struzik} and on the random events of the recording day.
On the other hand, dialogue between physicists and doctors seems to be inefficient:
physicists publish research results based on small test groups; 
doctors are waiting for follow-up studies using extended and homogeneous test groups, but
there seems to be no-one who has both necessary resources (funding, mathematical education, 
access to clinical data-bases), and motivation of doing such analysis.

\subsection*{Acknowledgments}
This study has been supported by the Estonian Science Foundation
grant No 4151.

\end{document}